\documentclass[
twocolumn,
secnumarabic,
amssymb, 
nobibnotes,
aps,prl,
superscriptaddress, 
nobalancelastpage,
longbibliography]{revtex4-2}

\usepackage{CJK}
\usepackage[tbtags]{amsmath}
\usepackage{graphicx}

\usepackage{tikz,xcolor,hyperref}
\definecolor{lime}{HTML}{A6CE39}
\DeclareRobustCommand{\orcidicon}{
	\begin{tikzpicture}
	\draw[lime, fill=lime] (0,0) 
	circle [radius=0.16] 
	node[white] {{\fontfamily{qag}\selectfont \tiny ID}};
	\draw[white, fill=white] (-0.0625,0.095) 
	circle [radius=0.007];
	\end{tikzpicture}
	\hspace{-2mm}
}
\foreach \x in {A, ..., Z}{%
	\expandafter\xdef\csname orcid\x\endcsname{\noexpand\href{https://orcid.org/\csname orcidauthor\x\endcsname}{\noexpand\orcidicon}}
}

\begin{document}
\begin{CJK*}{UTF8}{gbsn}

\title{The 3$\alpha$ correlations of ground and excited $0^+$ states of $^{12}\mathrm{C}$ within the microscopic cluster model}

\author{De-Ye~Tao (陶德晔)\orcidA{}}
\affiliation{Key Laboratory of Nuclear Physics and Ion-beam Application (MOE), Institute of Modern Physics, Fudan University, Shanghai 200433, China}

\author{Bo~Zhou (周波)\orcidB{}}
\email{zhou_bo@fudan.edu.cn}
\affiliation{Key Laboratory of Nuclear Physics and Ion-beam Application (MOE), Institute of Modern Physics, Fudan University, Shanghai 200433, China}
\affiliation{Shanghai Research Center for Theoretical Nuclear Physics, NSFC and Fudan University, Shanghai 200438, China}

\author{Yu-Gang~Ma (马余刚)\orcidC{}}
\email{mayugang@fudan.edu.cn}
\affiliation{Key Laboratory of Nuclear Physics and Ion-beam Application (MOE), Institute of Modern Physics, Fudan University, Shanghai 200433, China}
\affiliation{Shanghai Research Center for Theoretical Nuclear Physics, NSFC and Fudan University, Shanghai 200438, China}

\begin{abstract}
The cluster structures of the $0^+$ states in $^{12}\mathrm{C}$, including the ground state, the Hoyle state, and the recently identified $0_3^+$ and $0_4^+$ states, are analyzed to explore the cluster configurations and $3\alpha$ correlations without assuming the existence of $^8\mathrm{Be}$. 
In particular, the key quantity---two-cluster overlap amplitudes---is calculated for the $3\alpha$ clustering channels to reveal the essential features of these $0^+$ states.
The results clearly show the distinction between the compact structure of the ground state and the gas-like structures of the excited $0^+$ states.
The Hoyle state exhibits the expected gas-like dominant ($0S$) configuration, while the $0_3^+$ state shows a more extended $3\alpha$ clustering structure, which can be viewed as a breathing-like excitation of the Hoyle state, with an additional nodal structure.
The $0_4^+$ state is found to have a mixed configuration, featuring a bent-arm-like structure in the $S\otimes S$ channel and an enhanced $2\alpha$ correlation in the $D\otimes D$ channel.
\end{abstract}

\maketitle

\section{Introduction}
\label{intro}

Understanding the clustering structure of light nuclei has long been one of the most important issues in nuclear physics ~\cite{freer2018RMP035004,freer2014PPNP1,zhou2013PRL262501,enyo2007PTP655,tohsaki2001PRL192501,funaki2015PPNP78,chiba2017PTEP053D01,enyo2014PTEP073D02,ye2025NRP21,YeYL}, frequently applied in many studies of nuclear structure~\cite{HeWB,wang2022PRC034616,MaWH,MaCW,MaWH2,Natowitz}, astrophysical nucleosynthesis~\cite{dellaquila2024SR18958,adams2022AP102731,jin2020N57,Qin}, and heavy-ion collision~\cite{shen2024PRC064611,shi2021NST66,CaoYT}.
Extensive discussion on clustering effects in relativistic heavy-ion collisions is also under debate \cite{MaYG_book,Broniowski,ZhangS,LiYA,wang2022PLB137198,MaYG1,Liu,Prasad,MaYG2}. 
$^{12}\mathrm{C}$ as a typical clustering nucleus, is of special interest for unveiling the clustering structure~\cite{freer2018RMP035004}. 
In states above the $3\alpha$ break-up threshold, the structure and correlation of three $\alpha$ clusters in $^{12}\mathrm{C}$ are particularly intriguing~\cite{zhou2019PRC051303R,zhou2014PTEP101D01}.
Theoretically, the $3\alpha$ clustering in various states of $^{12}\mathrm{C}$, especially the well-known Hoyle state ($0_2^+$)~\cite{freer2014PPNP1} as a touchstone, has been investigated in the no-core Monte-Carlo shell model (MCSM)~\cite{otsuka2022NC2234}, the nuclear lattice effective field theory (NLEFT)~\cite{shen2023NC2777}, the antisymmetrized molecular dynamics (AMD)~\cite{enyo2007PTP655,enyo1998PRL5291}, the fermionic molecular dynamics (FMD)~\cite{chernykh2007PRL032501}, and analyses employing the Tohsaki-Horiuchi-Schuck-R\"opke (THSR) wave function~\cite{tohsaki2001PRL192501,zhou2016PRC044319,funaki2016PRC024344,funaki2015PRC021302}. 

In nuclear clustering physics, the cluster configuration in $^{12}\mathrm{C}$ has been extensively analyzed in numerous studies.
Within the cluster-model framework~\cite{funaki2015PPNP78}, the reduced-width amplitude (RWA), defined as the one-cluster overlap amplitude between the nuclear wave function and cluster wave functions, provides important insights into the structure and correlation of two-body clustering~\cite{tao2024,chiba2017PTEP053D01,enyo2014PTEP073D02}.
By calculating the $^8\mathrm{Be}+\alpha$ RWA, the $3\alpha$ configurations of $^{12}\mathrm{C}$ can be inferred, given that the ground state of $^8\mathrm{Be}$ is a well-developed $\alpha$-$\alpha$ resonance state with a $0S$ relative-motion orbit. 
For instance, the extended-tail and zero-node characteristics of $^{8}\mathrm{Be}(\mathrm{g.s.})+\alpha$ RWA for the Hoyle state suggest that the three $\alpha$ clusters equivalently occupy the lowest $0S$ orbit, forming a weakly-coupled gas-like structure~\cite{uegaki1979PTP1621}, which is consistent with the Bose-Einstein condensation (BEC) picture proposed in the THSR analysis~\cite{tohsaki2001PRL192501,zhou2019FoP14401,zhou2023NC8206,Kawa}. 
Furthermore, using the RWA, analogous $4\alpha$ and $5\alpha$ BEC states have been identified in $^{16}\mathrm{O}$~\cite{funaki2008PRL082502,funaki2010PRC024312} and $^{20}\mathrm{Ne}$~\cite{zhou2023NC8206}, respectively.
Apart from the ground and Hoyle states, two additional $0^+$ states recently observed just above the Hoyle state~\cite{itoh2011PRC054308,kurokawa2005PRC021301} have also been analyzed by calculating the $^{8}\mathrm{Be}+\alpha$ RWA.
It was found that the $0_3^+$ state likely represents a breathing-like excitation of the Hoyle state~\cite{zhou2016PRC044319,li2022PLB136928}, whereas the $0_4^+$ state is more likely to have a linear-chain or bent-arm structure~\cite{funaki2015PRC021302,funaki2016PRC024344}.

Many important features of clustering in $^{12}\mathrm{C}$ have been obtained by analyzing the $^{8}\mathrm{Be}+\alpha$ RWA.
However, this quantity essentially serves as a two-body clustering indicator.
To fully understand the nature of various $0^+$ states, it is critical to clarify the more fundamental $3\alpha$ correlations.
Consequently, it is highly desirable to extend the clustering analysis from the two-body $^{8}\mathrm{Be}+\alpha$ configuration to the three-body $\alpha+\alpha+\alpha$ configuration. 
Previously, to demonstrate $3\alpha$ correlations, the two-body density distribution of $\alpha$ clusters in some states of $^{12}\mathrm{C}$ has been presented in recent phenomenological structure studies~\cite{moriya2023EPJA197,moriya2023EPJA37} and some reaction studies~\cite{nguyen2013PRC054615,ishikawa2014PRC061604}.

Microscopically, the wave functions of several states of $^{12}\mathrm{C}$ were obtained over 50 years ago~\cite{uegaki1979PTP1621,kamimura1981NPA456,imai2019PRC064327}.
In this context, clarifying the evolution of the $3\alpha$ cluster structure as the excitation energy increases remains a central problem in clustering studies~\cite{zhou2019FoP14401}.
To precisely illustrate the $3\alpha$ structure and correlation based on microscopic wave functions of $^{12}\mathrm{C}$, the calculation method of RWA can be extended to obtain the two-cluster overlap amplitude (TCOA)~\cite{tao2024,descouvemont2023PRC014312,kimura2017PRC034331,kobayashi2016PRC024310}. 
This article aims to explore, using the microscopic cluster model, the detailed $3\alpha$ structure and correlation features of the Hoyle and $0_{3,4}^+$ states in $^{12}\mathrm{C}$.

\section{Two-cluster overlap amplitude}
\label{model}

To describe $^{12}\mathrm{C}$, the microscopic wave functions of the ground and excited states are obtained using the generator coordinate method (GCM)~\cite{hill1953PR1102,griffin1957PR311,horiuchi1977PoTPS90} as
\begin{equation}
    \Psi^{J\pi}_M = \sum_{i,K}c_{i,K} P_{MK}^{J\pi}\Phi^\mathrm{B}(\{\boldsymbol R\}_i),
\end{equation}
where the basis wave functions are adopted as Brink wave functions~\cite{brink1966}, specified by generator coordinates $\{\boldsymbol{R}\}=\{\boldsymbol{R}_1,\boldsymbol{R}_2,\boldsymbol{R}_3\}$. 
Before superposition, the basis wave functions are projected onto the eigenstates of angular momentum $J$ and parity $\pi$.
The coefficients $\{c_{i,K}\}$  are determined by solving the Hill-Wheeler equation. 

\begin{figure}
\includegraphics[width=0.6\linewidth]{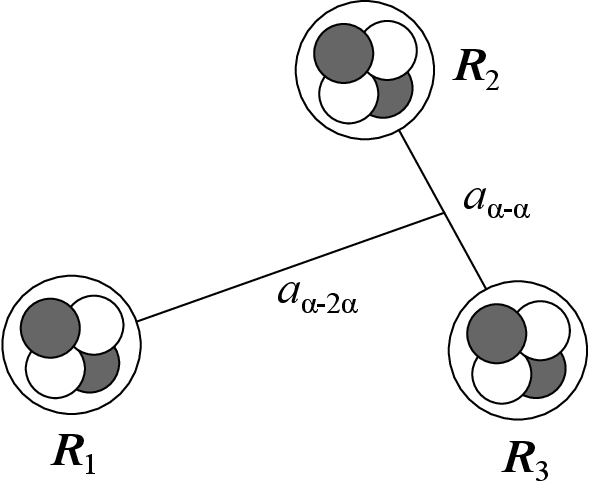}
\caption{\label{fig:coord} The illustration diagram for $\alpha+\alpha+\alpha$ clustering structure in $^{12}\mathrm{C}$ with the generator coordinates and $a_{\alpha\text{-}2\alpha}$, $a_{\alpha\text{-}\alpha}$ parameters.}
\end{figure}

To further analyze the three-cluster structure, the $3\alpha$ TCOA is defined as
\begin{widetext}
\begin{equation}
\begin{aligned}
    \mathcal{Y}_{L_{\alpha\text{-}2\alpha}L_{\alpha\text{-}\alpha}L}^{J\pi}&(a_{\alpha\text{-}2\alpha},a_{\alpha\text{-}\alpha})=\sqrt{\frac{12!}{4!4!4!}} \cr 
    &\times\left<\frac{\delta(r_{\alpha\text{-}2\alpha}-a_{\alpha\text{-}2\alpha})\delta(r_{\alpha\text{-}\alpha}-a_{\alpha\text{-}\alpha})}{r_{\alpha\text{-}2\alpha}^2r_{\alpha\text{-}\alpha}^2}\left[\left[Y_{L_{\alpha\text{-}2\alpha}}(\hat{r}_{\alpha\text{-}2\alpha})\otimes Y_{L_{\alpha\text{-}\alpha}}(\hat{r}_{\alpha\text{-}\alpha})\right]_L\otimes\left[\Phi_{\alpha}\otimes\Phi_{\alpha}\otimes\Phi_{\alpha}\right]_0\right]_{JM}\middle|\Psi_{M}^{J\pi}\right> ,
\end{aligned}
\end{equation}
\end{widetext}
with the channel configuration as illustrated in Fig.~\ref{fig:coord}.
$a_{\alpha\text{-}2\alpha}$ represents the distance between one $\alpha$ cluster and the center of mass (c.o.m.) of the other two $\alpha$ clusters. 
$a_{\alpha\text{-}\alpha}$ denotes the distance between the latter two $\alpha$ clusters.
$L_{\alpha\text{-}2\alpha}$ and $L_{\alpha\text{-}\alpha}$ denote the corresponding orbital angular momenta, while $L$ is the result of their coupling.
$\Phi_{\alpha}$ is the reference wave function of the $\alpha$ cluster.
For describing the three-body motion between $\alpha$ clusters, the dynamical relative-motion coordinates $\boldsymbol{r}_{\alpha\text{-}2\alpha}$ and $\boldsymbol{r}_{\alpha\text{-}\alpha}$ are defined as 
\begin{equation}
\begin{aligned}
    \boldsymbol{r}_{\alpha\text{-}2\alpha}&=\boldsymbol{X}_1-\frac{\boldsymbol{X}_2+\boldsymbol{X}_3}{2},  \\
    \boldsymbol{r}_{\alpha\text{-}\alpha}&=\boldsymbol{X}_2-\boldsymbol{X}_3,
\end{aligned}
\label{coord_jacobi}
\end{equation}
where $\boldsymbol{X}_i$ is the c.o.m.\ of the physical coordinates of the $i$th $\alpha$ cluster.
The overall component of $3\alpha$ clustering in a state of $^{12}\mathrm{C}$ can be evaluated using the three-cluster spectroscopic factor~(SF), defined as the integral of the squared norm of the TCOA for the coordinates $a_{\alpha\text{-}\alpha}$ and $a_{\alpha\text{-}2\alpha}$: 
\begin{equation}
    S_{3\alpha}^2=\int_0^\infty \int_0^\infty \left|a_{\alpha\text{-}2\alpha}a_{\alpha\text{-}\alpha}
    \mathcal{Y}_c^{J\pi}(a_{\alpha\text{-}2\alpha},a_{\alpha\text{-}\alpha})\right|^2 da_{\alpha\text{-}2\alpha}da_{\alpha\text{-}\alpha} .
\label{s_factor}
\end{equation}

\section{Results and Discussion}

Performing GCM calculations, we obtained the $0^+$ states of $^{12}\mathrm{C}$.
The nucleon-nucleon interaction is taken as the modified Volkov No.~2 potential~\cite{volkov1965NP33}, with parameters adopted the same as Ref.~\cite{kamimura1981NPA456}.
The binding energy of the ground state is calculated to be $89.65~\mathrm{MeV}$.
The Hoyle state is obtained to have an excitation energy of $7.82~\mathrm{MeV}$, which is very close to the $3\alpha$ break-up threshold.
The recently observed higher excited states, $0_3^+$ and $0_4^+$, are obtained in our calculation at $10.43~\mathrm{MeV}$ and $11.63~\mathrm{MeV}$, respectively.
Our results are in good agreement with previous REM~\cite{imai2019PRC064327}, THSR~\cite{zhou2016PRC044319} and RGM~\cite{kamimura1981NPA456} results, using identical interaction parameters.

\begin{figure*}
\includegraphics[width=\linewidth]{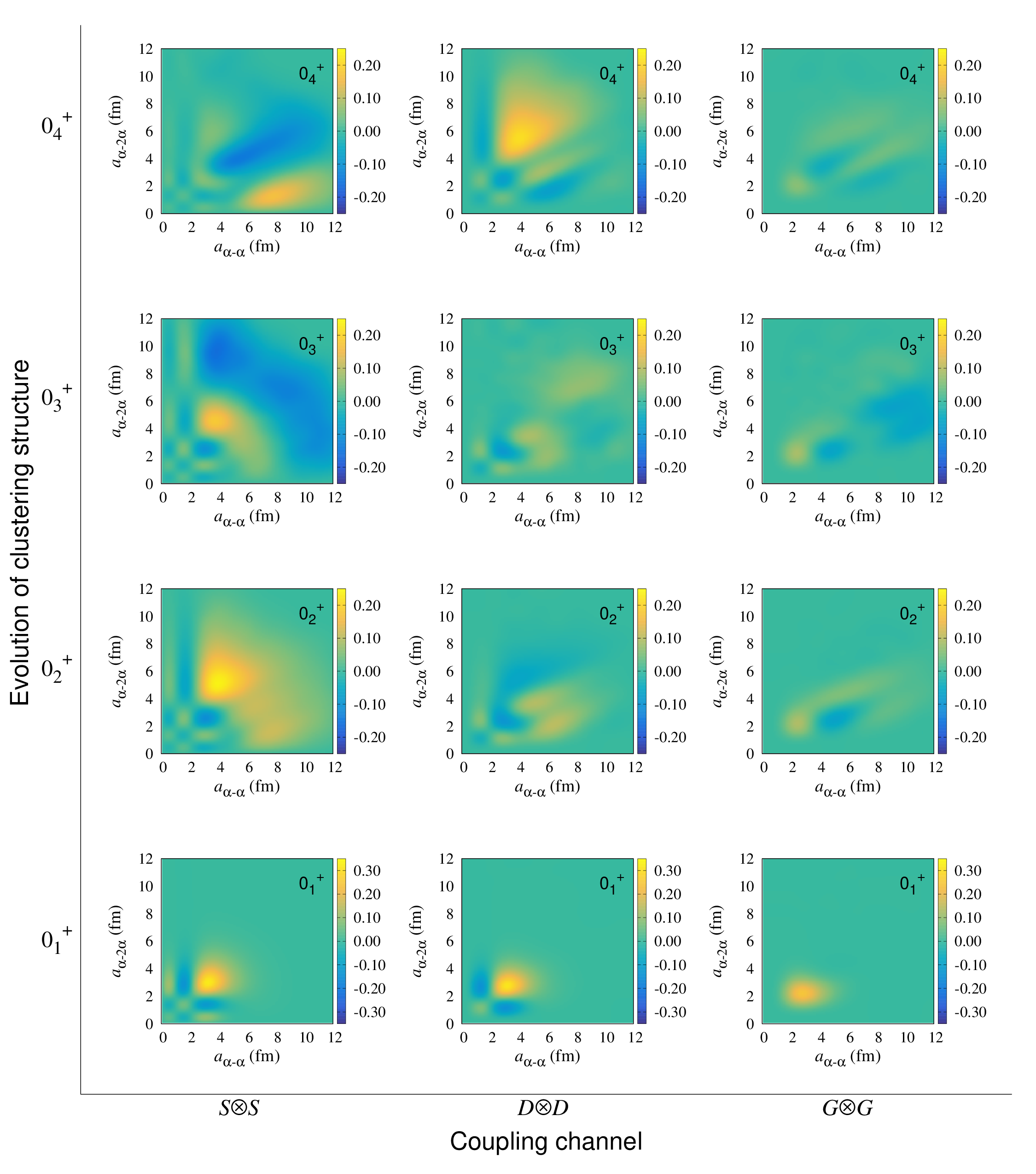}
\caption{\label{fig:tcoa} The calculated TCOA for the $0_1^+$, $0_2^+$, $0_3^+$, and $0_4^+$ states of $^{12}\mathrm{C}$.
The angular momentum coupling channel is denoted as $L_{\alpha\text{-}2\alpha}\otimes L_{\alpha\text{-}\alpha}$.}
\end{figure*}

\begin{figure}
\includegraphics[width=\linewidth]{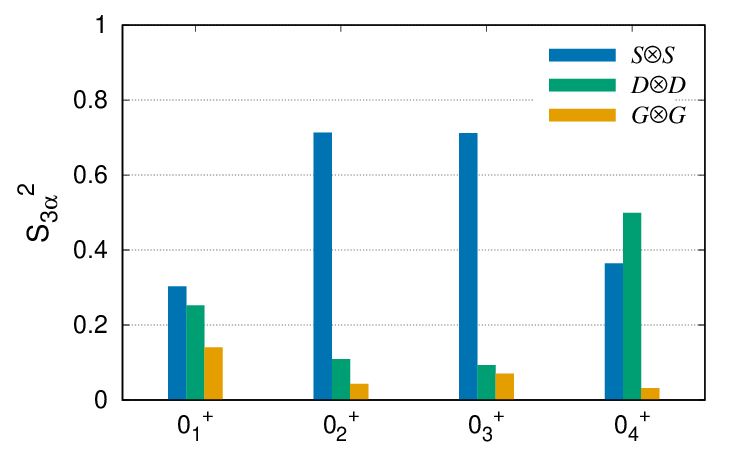}
\caption{\label{fig:sf} The calculated SF of $3\alpha$ clustering in the $0_1^+$, $0_2^+$, $0_3^+$, and $0_4^+$ states in $^{12}\mathrm{C}$.}
\end{figure}

The TCOA results for the ground state and three excited $0^+$ states of $^{12}\mathrm{C}$ are shown in Fig.~\ref{fig:tcoa}.
The amplitude distributions in different coupling channels show clear distinctions among these four $0^+$ states.
The ground state have compact configurations in all three channels of $S\otimes S$, $D\otimes D$, and $G\otimes G$.
The concentrated peaks for these channels are located at $(a_{\alpha\text{-}\alpha}=3.3~\mathrm{fm}, a_{\alpha\text{-}2\alpha}=3.0~\mathrm{fm})$, $(a_{\alpha\text{-}\alpha}=3.2~\mathrm{fm}, a_{\alpha\text{-}2\alpha}=2.8~\mathrm{fm})$, and $(a_{\alpha\text{-}\alpha}=2.8~\mathrm{fm}, a_{\alpha\text{-}2\alpha}=2.2~\mathrm{fm})$, respectively, suggesting a near-equilateral triangular arrangement of the three $\alpha$ clusters.
In the small-range areas, nodal lines appear due to non-negligible Pauli forbidden states, similar to nodes emerging in the RWA when the distance between clusters is small. 
The number of nodal lines is closely related to the corresponding orbital angular momentum.
In the $S\otimes S$ channel, each dimension has two nodal lines.
For $D\otimes D$, each has one.
The $G\otimes G$ amplitude has no nodal line.
It should be noted that the small-range nodal behaviour of the amplitudes of the ground state is nearly symmetrical for $a_{\alpha\text{-}\alpha}$ and $a_{\alpha\text{-}2\alpha}$, i.e., the relative motion between two $\alpha$ clusters when the third $\alpha$ is near their midpoint, and between one $\alpha$ and two strongly correlated $\alpha$ clusters~\cite{zhou2014PTEP101D01}.
The $S_{3\alpha}^2$ for the $S\otimes S$, $D\otimes D$, and $G\otimes G$ channels are calculated to be $0.303$, $0.252$, and $0.140$, respectively, as shown in Fig~\ref{fig:sf}.
The mixture of these competing channels with analogous patterns and the monotonic decrease in $S_{3\alpha}^2$ values with increasing orbital angular momentum are closely related to a shell-model-like structure.

In contrast to the compact ground state, the Hoyle state displays an extended and dilute amplitude, with the $S\otimes S$ channel being dominant.
This feature is consistent with previous studies, which suggest that the Hoyle state possesses a gas-like structure, regarded as a kind of BEC phenomenon for $\alpha$ clusters~\cite{tohsaki2001PRL192501}.
Nevertheless, nodal behaviour of the amplitudes persists when two of the $\alpha$ clusters are close to each other ($a_{\alpha\text{-}\alpha}<2.5~\mathrm{fm}$).
In this case, the fermionic nature of nucleons plays a significant role.
Note that the highest peak appears around the position of ($a_{\alpha\text{-}\alpha}=4.0~\mathrm{fm},a_{\alpha\text{-}2\alpha}=5.1~\mathrm{fm}$).
This coordinate corresponds to the structure where two $\alpha$ clusters are slightly closer to each other than to the third one, implying some $2\alpha$ correlation~\cite{zhou2014PTEP101D01} in this state, as pointed out in AMD and FMD analyses~\cite{enyo2007PTP655,chernykh2007PRL032501}.
This result is also consistent with the enhanced $^{8}\mathrm{Be}(\mathrm{g.s.})+\alpha$ RWA in previous studies~\cite{uegaki1979PTP1621,funaki2015PRC021302,zhou2016PRC044319}.
Besides the highest peak, two lower and narrower peaks exist in the area where $a_{\alpha\text{-}2\alpha}$ is smaller than $a_{\alpha\text{-}\alpha}$, corresponding to bent-arm-like or linear-chain-like structures, which have been obtained in MCSM and NLEFT calculations~\cite{otsuka2022NC2234,shen2023NC2777}. 
Additionally, the counterparts of these narrower peaks also emerge in the $D\otimes D$ channel, but with insignificant amplitudes.
The third $0^+$ state has been demonstrated to be a breathing-like excitation of the Hoyle state through RWA analyses~\cite{funaki2015PRC021302,zhou2016PRC044319}, supported by subsequent observations~\cite{li2022PLB136928}.
From the calculated TCOA, it is evident that, analogous to the Hoyle states, the $0_3^+$ state also exhibits an extended and dilute gas-like structure.
One distinction between the $0_3^+$ state and the Hoyle state is the additional nodal line that appears at relatively larger distances between clusters, around $8~\mathrm{fm}$. 
The similarity in $S_{3\alpha}^2$ between the $0_2^+$ and $0_3^+$ states is also notable. 
In Fig.~\ref{fig:sf}, concentrations of clustering components in the $S\otimes S$ channel for the $0_2^+$ and $0_3^+$ states are clearly shown.
Interestingly, the distributions of clustering components across different channels for these two states are also similar.
These results further support the conjecture that the $0_3^+$ state is an excitation of the Hoyle state with a breathing-like feature.
Such a notable similarity in clustering components was not observed when evaluating the SF of $^8\mathrm{Be}+\alpha$ configurations, which resulted in a much lower value in the $[^8\mathrm{Be}(\mathrm{g.s.})\otimes\alpha]\otimes S$ channel for the $0_3^+$ state compared to the Hoyle state~\cite{funaki2015PRC021302}. 
This can be attributed to the higher excitation energy of the $0_3^+$ state, which makes it difficult to maintain the ground state of $^8\mathrm{Be}$.
From this perspective, releasing the freedom of $3\alpha$ motion is significant for gaining a deeper understanding of the clustering structures and correlations of $^{12}\mathrm{C}$.

The structure configurations of the $0_4^+$ state show different features in the channels of $S\otimes S$ and $D\otimes D$, both of which have significant amplitudes.
In the $S\otimes S$ channel, two narrow peak regions appear where $a_{\alpha\text{-}2\alpha}<a_{\alpha\text{-}\alpha}$, indicating the bent-arm-like or linear-chain-like structures. 
This structural feature is consistent with recent THSR analyses~\cite{zhou2016PRC044319,funaki2016PRC024344}.
Note that for this channel, although definite shapes can be inferred, the amplitude distribution extends in two narrow areas, which should be considered a ``gas-like'' structure rather than a rigid-body structure.
Such an extremely deformed gas-like structure may correspond to the recently proposed one-dimensional $3\alpha$ condensation, as discussed in Ref.~\cite{suhara2014PRL062501}.
On the other hand, the amplitude for the $D\otimes D$ channel shows non-negligible values mainly in the area where $a_{\alpha\text{-}2\alpha}$ is larger than $a_{\alpha\text{-}\alpha}$, indicating two $\alpha$ clusters in closer proximity.
In this area, the amplitude distribution is also extended but in a relatively wide region, similar to the Hoyle state.
In terms of these characteristics, the $0_4^+$ state was actually obtained in previous GCM~\cite{uegaki1979PTP1621}, AMD~\cite{enyo2007PTP655,enyo1998PRL5291}, and FMD~\cite{chernykh2007PRL032501} calculations many years ago, but was recognized as the third $0^+$ state.
The calculated $S_{3\alpha}^2$ for the $0_4^+$ state in the $S\otimes S$, $D\otimes D$, and $G\otimes G$ channels are shown in Fig.~\ref{fig:sf}.
It can be clearly seen that the clustering configuration is partially attributed to the first two channels, with the $D\otimes D$ component slightly larger than the $S\otimes S$ component.
Corresponding to $S_{3\alpha}^2$ for the $S\otimes S$ channel, the SF of the two-body $[^8\mathrm{Be}(\mathrm{g.s.})\otimes\alpha]\otimes S$ configuration also underestimated the $3\alpha$ clustering component in this channel.
Thus, the structure of the $0_4^+$ state can be concluded to be a mixture of two gas-like configurations of bent-arm-like $3\alpha$ in an $S$-wave orbit and $2\alpha$ correlation in a $D$-wave orbit.

The evolution of $3\alpha$ clustering in $0^+$ states of $^{12}\mathrm{C}$ can be well interpreted by the calculated TCOA.
Governed by the Pauli exclusion principle, $\alpha$ clusters in the ground state are geometrically confined to a near-equilateral triangle with a side length of about $3~\mathrm{fm}$.
Excitation to the energy near the $3\alpha$ break-up threshold gives rise to evident $3\alpha$ clustering in the Hoyle state, the first excited $0^+$ state, where the bosonic characteristics of the $\alpha$ clusters influence the nuclear structure. 
In this state, the clusters can move in a large spatial extension with relatively weak Pauli repulsion. 
The $3\alpha$ motion is further excited as the energy increases.
Note that the $0^+$ states should be orthogonal to each other, and consequently, fundamentally different characteristics must be ensured among them.
Sharing similar clustering components with the Hoyle state, the $3\alpha$ motion of the $0_3^+$ state is excited radially to a breathing-like mode to ensure the orthogonality.
The $0_4^+$ state, on the other hand, is excited to an extremely deformed bent-arm-like configuration in the $S\otimes S$ channel, mixed with comparable components of $2\alpha$-correlated configurations in the $D\otimes D$ channel.
Remarkably, the gas-like feature of $3\alpha$ clustering structures is found not only in the Hoyle state and its breathing-like excitation but also in the $0_4^+$ state, for both the $S\otimes S$ and $D\otimes D$ channels.
Moreover, given the non-negligible $2\alpha$ correlations emerging in these $0^+$ states, exploring the rotational band configurations involving higher angular-momentum states will be addressed in our forthcoming paper.

\section{Summary}

In summary, the $3\alpha$ clustering of $0^+$ states in $^{12}\mathrm{C}$ is microscopically analyzed through the calculation of TCOA.
It is shown clearly that the ground state of $^{12}\mathrm{C}$ has a compact structure, tending to form a near-equilateral triangle, while the excited $0^+$ states exhibit gas-like configurations. 
Both the Hoyle state and the $0_3^+$ state have predominant amplitudes in the $S\otimes S$ channel, supporting the idea that the $0_3^+$ state can be considered a breathing-like excitation of the Hoyle state. 
The $0_4^+$ state is found to have a gas-like, extremely prolate-deformed structure in the $S\otimes S$ channel, which may be associated with one-dimensional $3\alpha$ condensation, and a loosely bound $2\alpha$-correlated structure in the $D\otimes D$ channel.
It should also be noted that the overlap amplitudes bridge the nuclear structure and reaction analyses, serving as important input data to obtain cross sections~\cite{yoshida2019PRC044601,yoshida2018PRC024614,lyu2018PRC044612}.
In the future, the calculated TCOA may provide more microscopic structural information and improve the calculations of two-cluster (nucleon) knock-out reactions.

\begin{acknowledgments}
The authors would like to thank Prof.~M.~Kimura, Prof.~K.~Kato, Prof.~T.~Yamada, and Prof.~Y.~Funaki for their useful discussions. This work is supported by the National Key R$\&$D Program of China (2023YFA1606701). This work was supported in part by the National Natural Science Foundation of China under contract Nos. 12175042, 11890710, 11890714, 12047514, 12147101, and 12347106, Guangdong Major Project of Basic and Applied Basic Research No. 2020B0301030008, and China National Key R$\&$D Program No. 2022YFA1602402. This work was partially supported by the 111 Project.
\end{acknowledgments}

\end{CJK*}
\end{document}